# Nanoscale control of molecular self-assembly induced by plasmonic hot-electron dynamics

Sabrina Simoncelli[1‡*], Yi Li[1‡], Emiliano Cortés[1] and Stefan A. Maier[1]

[1]The Blackett Laboratory, Department of Physics, Imperial College London, London SW7 2AZ, United Kingdom

‡These authors contributed equally to this work
* Corresponding author: s.simoncelli@imperial.ac.uk

**Abstract:**

Self-assembly processes allow us to design and create complex nanostructures using molecules as building blocks and surfaces as scaffolds. This autonomous driven construction is possible due to a complex thermodynamic balance of molecule-surface interactions. As such, nanoscale guidance and control over this process is hard to achieve. Here we use the highly localized light-to-chemical-energy conversion of plasmonic materials to spatially cleave Au-S bonds on pre-determined locations within a single nanoparticle, enabling unprecedented control over this archetypal system for molecular self-assembly. Our method offers nanoscale precision and high-throughput light-induced tailoring of the surface chemistry of individual and packed nano-sized metallic structures by simply varying wavelength and polarization of the incident light. Assisted by single-molecule super-resolution fluorescence microscopy, we image, quantify and shed light onto the plasmon-induced desorption mechanism. We show that plasmon-induced photo-desorption enables sub-diffraction multiplexing, selective positioning of nanomaterials and even sorting and purification of colloidal nanoparticles.

**Introduction:**

Self-driven assembly of thiolated molecules onto metals offers a direct route to tailor the surface chemistry and functionalities of nanomaterials, cementing the bottom-up hierarchal construction of complex nanosystems.[1,2] The robust metal-sulfur bond formation drives the self-assembly process and the strength of van der Walls interactions between molecules stabilise and order the molecular layer, hindering their diffusion and turning them amenable to pattern surfaces properties by almost free election of their terminal group.[3–5] There is a wide variety of either bio-functional molecules (DNA, proteins, aminoacids) or inorganic molecules, complexes and polymers that intrinsically have or can be further modified with a thiol group. As such, thiols are the natural workhorse molecular linkers in various fields of nanoscience including molecular electronics, catalysis, sensing and medical therapy, among many others.[3,6] Given the technological importance of these surface chemistry processes, controlling them down to the nanometer scale can impact in various fields of current nanoscience and nanotechnology. At the moment, there are no simple and easy-to-implement strategies to '*in-situ*' control the self-assembly process down to the sub-nanoparticle level. Some advances on this direction, although limited in achieving sub-nanoparticle precision, include techniques such as micro-contact printing,[7] direct laser patterning[8] and electrochemical stripping[9]. These micro-scale pattering self-assembly techniques are deceptively simple; their spatial resolution is highly dependent upon fabrication of stamps and/or transfer conditions, the laser beam size and power or the size of the metallic micro-electrodes, respectively. In this regard, Mirkin et al. pioneered on achieving highly reproducible nanoscale manipulation (~ 50 nm) of alkanethiols self-assembly with the introduction of dip-pen nanolithography.[10] Dip-pen nanolithography exploits the capillary action between an atomic force microscope tip and a gold substrate to produce multi-molecule patterns, withal, it tends to be technically involved to implement and it is a time-consuming 'bottom up' process. Alternative methods to spatially control the surface chemistry of nano-sized structures and colloids could revolution our current abilities to produce multi-functional nano-materials.

It is well known that thiol-release from metal surfaces can be induced by temperature, light, voltage and chemical reagents.[3–5,11,12] However, none of these routes have proven specificity and sub-diffraction spatial control over this process. To this end, deep understanding of the reactivity of the Au-S bond under external stimuli could open interesting possibilities for tailoring functionalities and the manipulation of nanostructures. In particular, the enhanced absorption of photons in metal nanostructures in comparison to bulk metal surfaces and the subsequent excitation and non-radiative decay of surface plasmon resonances sets a new paradigm for light-into-chemical energy conversion.[13–16] Plasmon-enhanced photo-catalytic processes, such as CO oxidation, $H_2$ dissociation, nitro to amino conversion and water splitting, among others, have been successfully achieved.[17–19] Photo-release of thiolated single-stranded DNA molecules from gold nanoparticles was also reported.[11,12] Although a lot is known about plasmon-induced photochemical processes, the potential to engineer spatial and/or chemical selectivity has no yet been exploited. Here we take advantage of the plasmon-mediated light-induced cleavage of thiol molecules on the surfaces of gold nanostructures followed by ligand-replacement for 'in-situ' nano-localized patterning. We use super-resolution imaging to demonstrate the hot-electron driven nature of the process. The confined and tunable hot-electrons localization in plasmonic materials offers a new degree of freedom to tailor surface chemistry self-assembly. Notably, understanding plasmonic–photonic interactions allows for the predesign of specific functionalities on either sub-particle surfaces or nanosized elements of complex metallic nanoclusters. Our results open new perspectives for the design, fabrication, and applications of functional platforms in the nanoscale regime.

**Results:**

Figure 1a illustrates the concept of our simple yet arguably elegant plasmon-induced spatial functionalization approach. First, a monolayer of thiol reactive molecules is self-assembled on electron-beam patterned gold nanoantenna arrays. We use dark-field scattering spectroscopy and numerical simulations to determine the plasmon resonance wavelengths of each fabricated structure. Next, individual functionalized gold nanoantennas embedded in a microfluidic chamber are irradiated with a wavelength-tunable linearly polarized 180 fs pulsed laser. Due to the resonant energy transfer into electron cloud dipole oscillations, we would expect an enhanced optical absorption on the nanoantenna element(s) parallel to the excitation polarization. Therefore, only the thiol molecules localized on the resonantly excited region of the nanoantenna are expected to detach from the surface, leaving it now accessible for further functionalization with another type of molecules. Finally, a second step of incubation is performed on the irradiated gold nanoantenna to selectively bind a new type of thiol-reactive probe to the bare elements. Automation of this procedure is easily achieved by combining the use of a fluidic chamber, for rapid fluid exchange, and of a nanopositioning piezo stage to irradiate micro-sized arrays of nanostructures in a just a few seconds. As such, this approach can be extended to incorporate multiple functional groups with nano-sized precision on selective surfaces of plasmonic antennas by consecutive irradiation and incubation steps.

**m-PAINT super-resolution imaging of molecular targets on metallic surfaces**

To map the surface chemistry evolution of gold nanoantennas with sub-diffraction resolution, we adapt a point-localization-based super-resolution fluorescence technique, named DNA-PAINT (point accumulation for imaging in nanoscale topography).[20,21] DNA-PAINT employs transiently binding of short fluorescently labelled oligonucleotides ('imager strands') to complementary single-stranded DNA oligos ('docking strands') chemically coupled to a target object. Figure 1b shows the scheme of the super-resolution metallic DNA-PAINT (m-PAINT) imaging technique. First, docking strands, chemically coupled to a supplementary hydrophilic PEG unit in the form of hexaethylene glycol, are docked to the metallic surface through a mercaptohexyl linker (Figure 1b, i).[22] The introduction of such a PEG moiety optimises the separation distance between the fluorophores and the metallic surface to ~ 7 nm and as such, it overcomes the spatial fluorescence quenching effects on metals.[23,24] Additionally, we choose a backfilling molecule (6-mercapto-1-hexanol, MCH) to help maintaining the availability of the ssDNA docking molecules by blocking binding sites and thus preventing unspecific binding.[25] Finally, the density of docking strands on the surface of the gold nanoantennas are comprehensively optimised by tuning the assembly time, thiolated (hence docking) ssDNA concentration, and salt concentration. From reductive electrodesorption measurements we estimate a ratio of ~ 1:160 between ssDNA and MCH molecules (see Supplementary Section 6 and Supplementary Figure 1), which results in ~ 5 binding events per image frame. This number of events was sufficient to reconstruct in 20,000 frames an accurate representation of the nano-sized objects as well as the localization of different types of target molecules over the surface of the nanoantennas (Figure 1b, iv-ix). We evaluated the imaging performance of m-PAINT by visualization of a sub-diffraction six-leg gold nano-assembly structure (Figure 1b, iv). The individual super-resolved images show a clear increase in spatial resolution from the diffraction-limited representation (Figure 1b, v-viii). Comparing the SEM images and the super-resolved images of individual six-leg structures demonstrates that we can successfully resolve metallic nanostructures using m-PAINT. To account for the fact that more localization events render a better-defined nanoantenna image, we overlaid super-resolution images of 36 individual gold nanostructures (see Supplementary Section 9, Supplementary Figure 2 and Figure 1b, ix). The high-density super resolution image of the six-leg structure accurately represents the geometry of the fabricated one with a precision of ~ 20 nm (Figure 1b, iii).

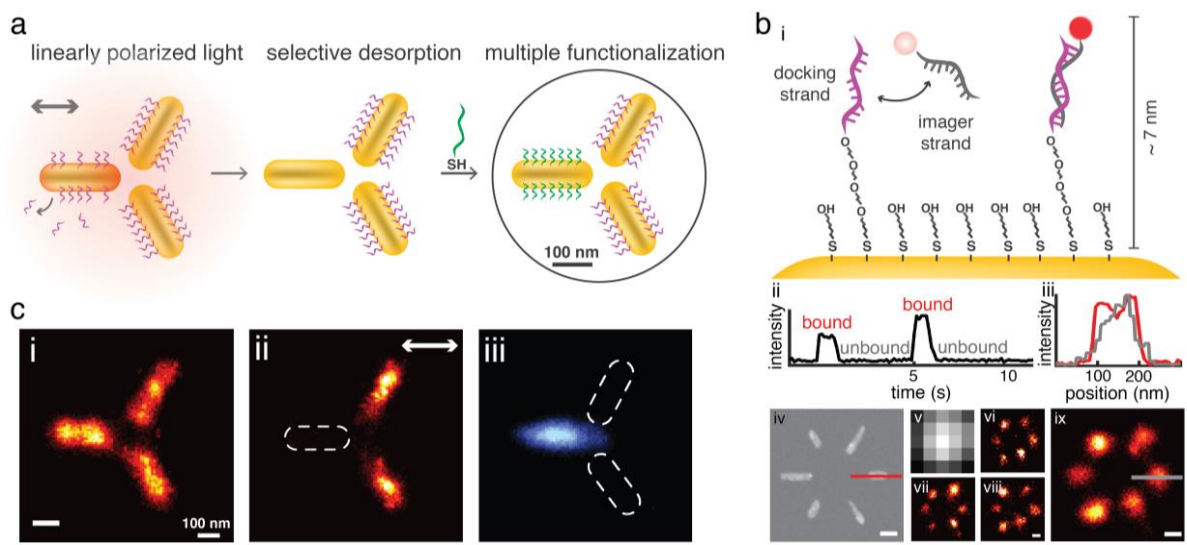

**Figure 1. Plasmon-selective surface chemistry modification. a,** Optical excitation with a linearly polarized femtosecond pulsed laser enables selective and nano-localized molecular desorption of thiol-functionalized Au nanostructures. The unprotected region is now accessible for re-functionalization with a new thiol-reactive probe. **b,** Super-resolution metallic DNA-PAINT imaging technique (i). Transient binding of fluorescently labelled DNA ('imager') strands to their complementary sequence ('docking') strands are detected as a switching between a dark (unbound) and a bright (bound) state (ii). SEM (iv), diffraction-limited (v) and super-resolved m-PAINT images (vi - viii) of individual Au six-leg nanostructures. Spatial intensity profile of the SEM (red) and the average super-resolved six-leg image (grey) along the horizontal short leg of the nanostructure (iii). Average super-resolution image (x) of 36 individual six-leg Au nanostructures (see Supplementary Figure 1). **c,** Average super-resolved m-PAINT images of 28 individual Au nanotrimers before (i) and after (ii, iii) irradiation with a 950 nm linearly polarized femtosecond pulsed laser (1.7 mJ/cm$^2$) and re-functionalization with the thiol-reactive docking strand P$_2$ (iii). Pseudo-colors red and blue for docking strands P$_1$ and P$_2$, respectively. The mean number of localizations in individual images is 1,608 ± 564, 776 ± 480 and 1,080 ± 597 for i to iii.

## Selection rules for controlling the nano-localized Au-S desorption

We first tested the performance of the plasmon-selective functionalization procedure on trimer gold nanorods. The wavelength of irradiation was chosen to excite the longitudinal plasmonic mode of the nanorods (Supplementary Figure 3). Figure 1c shows the m-PAINT super-resolution average images of nanotrimer arrays before (i) and after (ii) irradiation with a 950 nm linearly polarized femtosecond pulsed laser. After laser irradiation, we only observed DNA binding fluorescent events in the nanorods whose orientation is not parallel to the laser polarization. Note, however, that the spot size of the ultrafast laser was ~ 2.7 μm, measured via determining its point spread function (PSF), meaning that the whole nanotrimer structure is being irradiated. This indicates that only the organothiol molecules localized on the surface of the resonantly excited nanorod were selectively desorbed after irradiation (Figure 1c, ii). The reproducibility of the technique is further evidenced by the super-resolution images of 28 individual irradiated nanoantennas (Supplementary Figure 4). We achieved bi-functionalization in a sub-diffraction limited area by incubating the sample with a new type of thiol-labelled docking strand (P$_2$). The second round of super-resolution imaging (imaging strand P$_2$*) demonstrates that our plasmon-selective functionalization approach is a rapid method for tailoring the chemical surface of plasmonic nanostructures with sub-diffraction resolution (Figure 1c, iii). This experiment also confirms that the integrity of the excited nanorod is maintained after irradiation (i.e. gold antennas do not melt). We also demonstrated that when the trimer is resonantly excited, we can selectively desorb and re-functionalize thiol-labelled molecules from any desired rod of the nanostructure just by rotating the polarization of the incident light (Supplementary Figure 5).

We generalise this idea to complex gold nanoclusters, where on-demand patterning requires both wavelength and polarization selection rules of the incident light. Figure 2a shows the two building blocks for our demonstration nanoclusters, whose resonances are well separated at 835 nm and 1040 nm (Figure 2b). We first assess the specificity provided by excitation wavelength on gold nanoclusters composed of six nanorods. Figure 2g and 2h depict the metallic DNA-PAINT images of the six-leg nanostructures irradiated with two distinct wavelengths. The preferential desorption of the thiol labelled docking strand $P_1$ from the resonantly excited nanorods indicates that spectral differences provide an additional efficient way to tune the surface chemistry of packed nanoclusters, in addition to polarization as demonstrated in Figure 1. We then constructed a letter pattern 'ICL' (abbr. of Imperial College London), shown in Figure 2e, for multiplexing at two wavelengths and two polarizations in a structure where plasmonic near-field coupling may hinder the general use of our technique. Even though the plasmonic resonances between the two building blocks of the 'ICL' design couple (Supplementary Figure 7, vertical polarization), Figure 2i and 2j shows Au-S desorption with high selectivity. Thus, our experimental results indicate that near-field cross-talk among the nanorods does not contribute significantly to the general Au-S desorption mechanism.

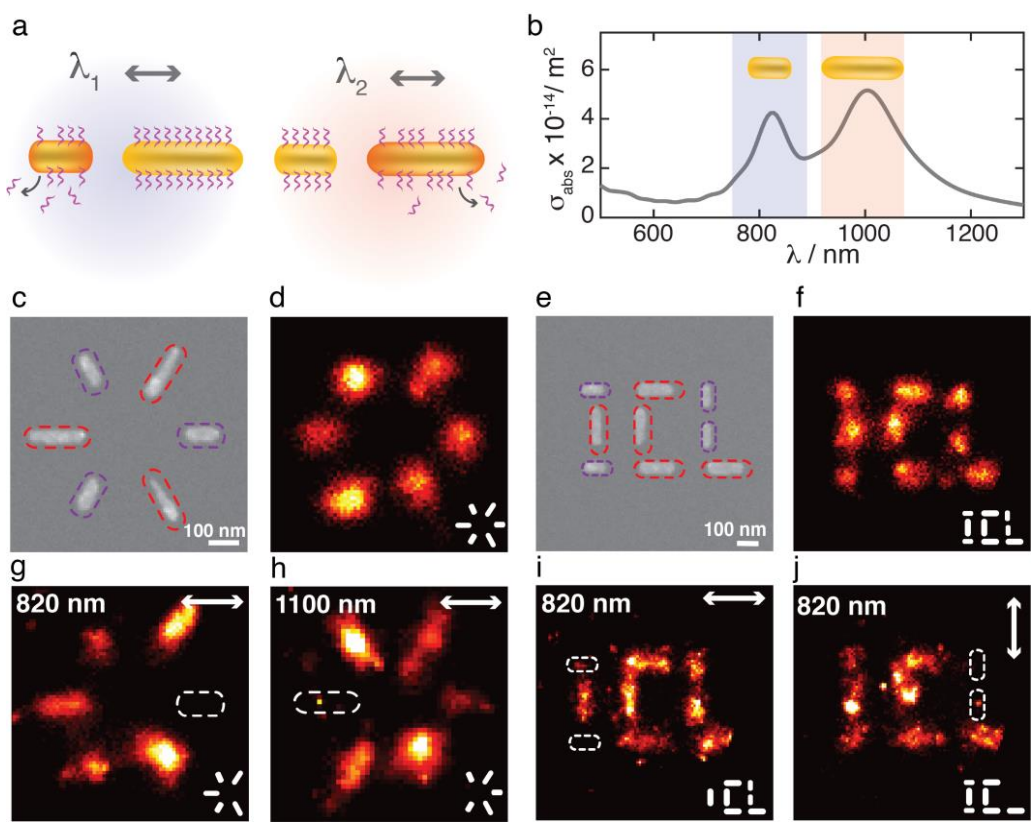

**Figure 2. Wavelength and polarization control of thiol-desorption. a,** The spectral differences between nanorods that share the same orientation allow tuning the surface chemistry of packed nanostructures. **b,** FDTD-simulated single-antenna absorption spectra of a two-leg Au nanostructure (dimensions: 170 x 50 nm, 110 x 45 nm, gap 120 nm). **c, e,** SEM image of a six-leg or a custom-design letter pattern ('ICL') Au nanostructure composed of two sizes of nanorods with dimensions of 110 x 45 nm (violet dashed line) and 170 x 50 nm or 180 x 40 nm (red dashed line) for the ICL and the six-leg nanostructure, respectively. **d, f, g-j,** Average super-resolved m-PAINT images of 35 individual Au nanostructures functionalized with docking strand $P_1$ before (**b, h**) and after (**e, f, k, l**) irradiation with a 820 nm (**g, h, i**) or a 1100 nm (**j**) linearly polarized femtosecond pulsed laser of 2.6 mJ/cm$^2$ (**g, h**) or 1.9 mJ/cm$^2$ (**i**) or 3.8 mJ/cm$^2$ (**j**) fluence. White arrows represent the polarization of the incident light. The mean number of localizations in individual images is 1,458 ± 332; 1,579 ± 351; 1931 ± 672; 1,580 ± 496; 970 ± 384 and 689 ± 263 for **d, f-j**, respectively.

## Spatial and temporal study of Au-S photo-desorption mechanism

In order to assess the extent of applicability of our plasmon-selective surface chemistry modification approach, we next investigated the Au-S desorption mechanism. Plasmon-induced Au-S photo-desorption can be triggered by one of the two general mechanisms facilitated by the dominant non-radiative plasmon decay channels: localized heating of the surrounding medium or 'hot'-carrier-driven chemical reactions.[11,12,17,26] Following femtosecond laser irradiation, surface plasmons decay into anisotropic distributions of hot electrons (or holes) within ~ 40 fs.[18,27–29] While these energetic electrons are localized in regions of dense resistive losses of the particle, they later (on a time scale of ~800 fs) redistribute their energy via electron-phonon scattering, which leads to a uniform lattice temperature distribution due to reasonable high thermal conductivity and longer response time (see Supplementary Section 12).[18,30] Therefore, selective control over the spatially localized chemistry of a single plasmonic nanoantenna can only be achieved prior to this electron-lattice thermalization.[19] To straightforwardly assess the origin of the gold-thiol desorption mechanism, we designed a branched plasmonic antenna, such as a 'L' shape letter, capable of differentiating between the two photo-driven processes (Figure 3a-c).

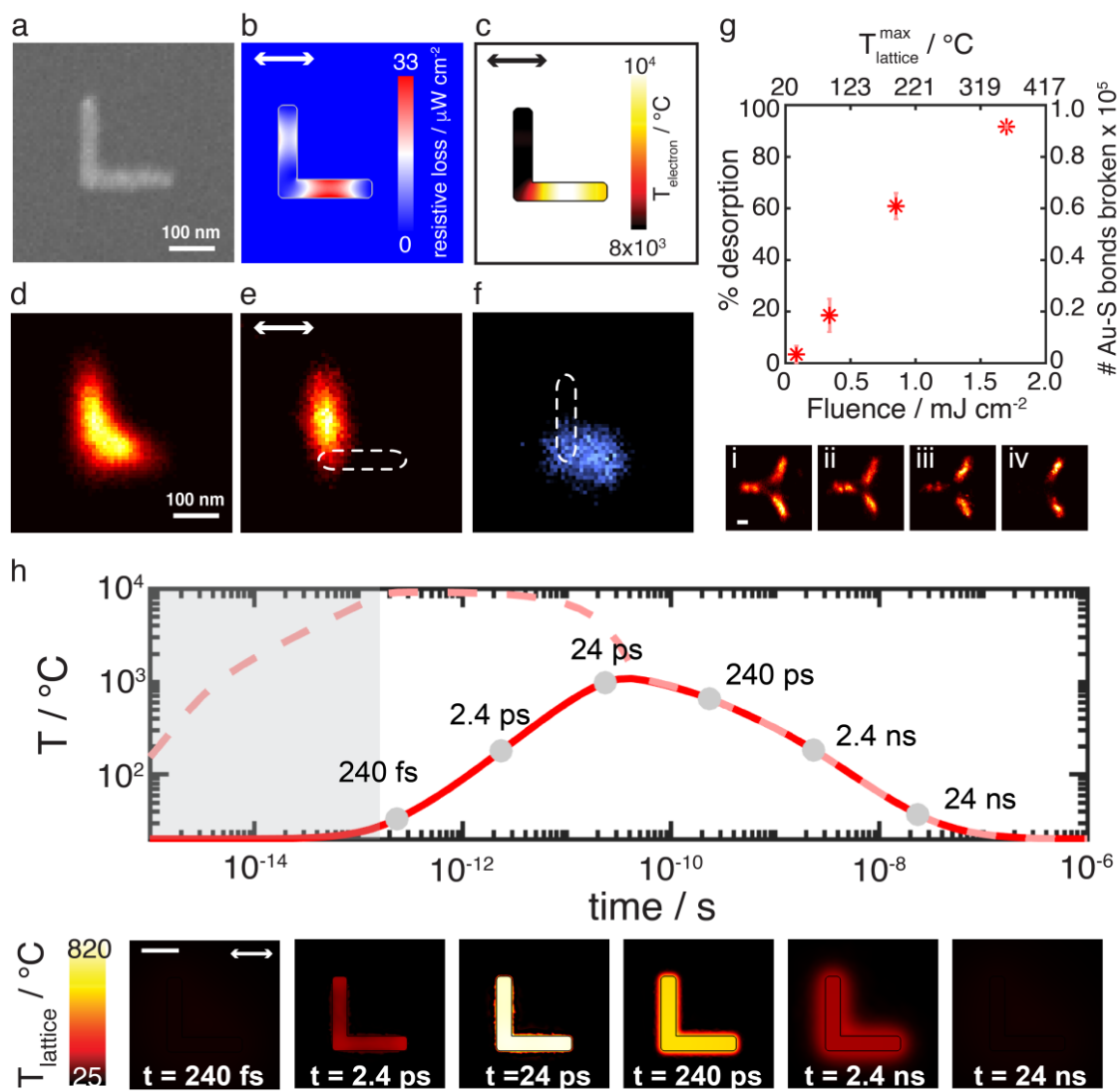

**Figure 3. Plasmon-selective Au-S desorption mechanism. a,** SEM image of an L-shape Au nanostructure. **b, c,** Resistive loss and electron temperature distribution at 240 fs delayed from the temporal start of a 180 fs laser pulse (950 nm, 5.3 mJ/cm$^2$). **d-f,** Average super-resolved m-PAINT images ($n$ = 31) before (**d**) and after (**e, f**) irradiation (950 nm, 5.3 mJ/cm$^2$) and re-functionalization (**f**) with the thiol-reactive docking strand P$_2$. The mean number of localizations is 2,016 ± 380, 701 ± 436 and 260 ± 202 for **d** to **f**. **g,** Power dependent Au-S desorption of docking strand P$_1$ from Au nanotrimers after femtosecond pulse irradiation (0°, 950 nm). Error bars correspond to the average of two independent data sets. i-iv, Average super-resolved m-PAINT images ($n$ = 31) of nanotrimers after laser irradiation of 0.08, 0.3, 0.6 and 1.7 mJ/cm$^2$ fluence, respectively. The mean number of localizations is 2,764 ± 804, 1,586 ± 621, 1,157 ± 463 and 776 ± 480 for i to iv. **h,** Simulated temporal evolution of the electron temperature (dashed line) and lattice temperature (full line) following femtosecond laser irradiation. Map distribution of the lattice temperature at different delays after the temporal start of the fs laser pulse.

Figure 3d-f shows the super-resolved fluorescence images of 'L' connected gold nanoantennas before and after irradiation. The spatially resolved images reveal that depending on the polarization of the incident laser only molecules localized either on the horizontal or on the vertical leg of the 'L' structure desorbed (Supplementary Figure 8). These results are in line with a hot-electron initiated process, in which asymmetric distribution of resistive losses induces localized reductive desorption of the thiolated molecules. Numerical simulations to map the lattice temperature and the conduction electrons density of the 'L' shaped nananoatenna – under femtosecond pulsed illumination – also support a hot-electron initiated process rather than a phonon driven reaction for the photo-desorption of thiol molecules (Figure 3h and Supplementary Figure 8). Notably, the intrinsic short lifetime of hot electrons offers unique confinement to tailor sub-particle surface chemistry (Figure 3e and 3f). Contrary to excited carriers, the uniform nanosecond lattice heating has no spatial resolution for thermally induced desorption in single nanoantennas. Furthermore, the kinetics for thermal induced thiol-release should be hindered on this timescale.[12]

Another evidence, which points to a hot-electron-driven-mechanism, is the linear dependence of the Au-S release on laser intensity (Figure 3g).[17] Contrary to a thermal activated process, where a Heaviside step function is expected once the desorption temperature is reached,[11] we observed a monotonic increase in the desorption percentage with increasing the laser power.[31] This is consistent with a hot-electron-initiated bond dissociation reaction, where the number of generated hot carriers has a positive relationship with the photon flux. Indeed, we estimate that 10 Au-S bonds are cleaved per pulse of light for a fluence of 1.7 mJ cm$^{-2}$, a similar figure to the one reported for hot-electron induced $H_2$ dissociation in Al-Pd nanoantennas.[32]

**Applications of plasmon-guided ligand replacement**

Finally, we show the versatility of the plasmon-selective functionalization procedure with two clear examples where self-induced capping removal in metal nanoparticles is used to: 1) guide the positioning of nano-objects at the nanoscale and 2) solution-based sorting and purification of metal nanoparticles. For the first example, we start by functionalizing the nanoantennas with a self-assembled monolayer of MCH molecules followed by targeted MCH photo-desorption. Ligand replacement allows the self-assembly of thiolated biotin-terminal linkers at specific locations of the nanostructure – i.e. only at the places where MCH was previously removed – which can further bind streptavidin coated gold nanoparticles. Figure 4b-e depict the preferential attachment of streptavidin coated 50 nm gold nanoparticles to one or two rods of the nanotrimer depending on the direction of the incident light in the initial removal process. Although further optimization is required to attain a more controllable nanoparticle binding, our results show the potential of using plasmon-selective desorption to enable hierarchical nanoparticles assembly. Figure 4f-h, on the other hand, shows a proof of concept that further explores the wavelength dependence of the Au-S bond cleavage in colloidal plasmonic antennas. Selective thiol displacement from a colloidal mix of 50 nm Au spherical nanoparticles (AuNPs) and 150 x 50nm Au nanorods (AuNRs) was achieved by illuminating the solution at the resonance wavelength of one of the constituents (i.e. 780 nm, AuNRs longitudinal plasmon resonance) at a laser power below the onset energy for melting and reshaping of single gold nanorods.[33] Both types of nanoparticles were initially coated with a self-assembly monolayer of 4-mercaptobenzoic acid (MBA). After the illumination a clear aggregation of the AuNRs takes place as a consequence of thiol displacement only for this type of particles, as shown in the extinction spectra and SEM images for the pre/post irradiated solutions. AuNPs remain unaltered, highlighting the highly wavelength-dependent specificity of the process. This proof of concept not only extends our findings to solution-based experiments but also demonstrates that in principle any thiolated molecule can be also selectively displaced, as the energy of the Au-S bond is nearly independent on the nature of the backbone chain.[3,34]

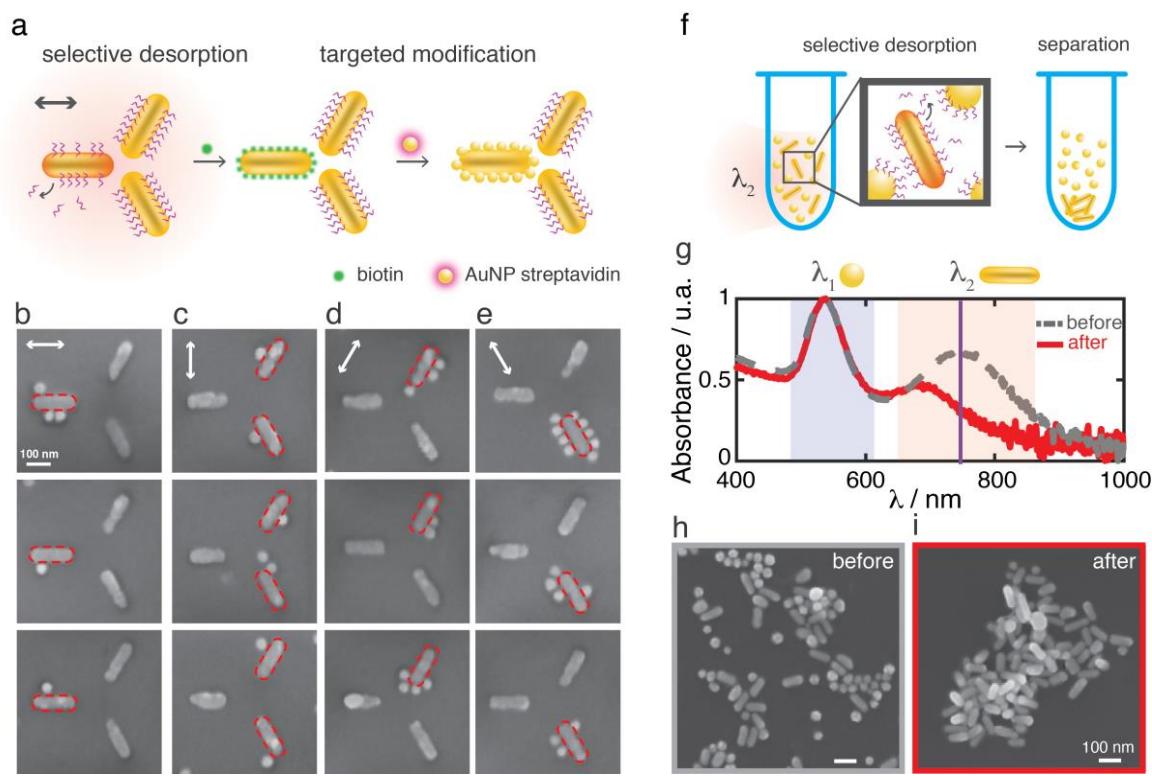

**Figure 4. Hierarchical assembly of metal nanostructures and solution-based sorting of metal nanoparticles. a,** Au nanostructures are first modified with a self-assembly monolayer of 6-mercapto-1-hexanol and then illuminated with a linearly polarized femtosecond laser at their local surface plasmon resonance wavelength (950 nm, 1.7 mJ/cm$^2$). The samples are rinsed and immediately functionalized with a biotin terminal group. Streptavidin coated Au nanoparticles (50 nm) are then left to react overnight with the bi-functionalized Au nanostructures. SEM images are acquired after several washing steps with PBS buffer. **b-e,** Representative SEM images of the 50 nm Au nanoparticles binding to the fabricated bi-functionalized gold nanostructures. White arrows represent the polarization of the incident light. **f,** Au nanoparticles and nanorods modified with a self-assembly monolayer of 6-mercapto-1-hexanol are illuminated with a linearly polarized femtosecond laser at the nanorod local surface plasmon resonance wavelength (750 nm, 8 mJ/cm$^2$). S-Au desorption from Au nanorods induces the aggregation of the nanorods in the solution. **g, h,** Absorbance spectra and representative SEM images of the 50 nm Au nanoparticles and the 50 x 150 nm Au nanorods solution mixture before and after laser irradiation.

## Conclusions:

In summary, we have described a simple, rapid and robust nanofabrication approach to control, with sub-diffraction precision, the surface chemistry of individual and clustered metallic nanoantennas. While continuous wavelength illumination of plasmonic nanostructures can induce chemical changes in the terminal groups of self-assembled monolayers, pulsed-illumination enables cleavage of their anchor group.[11,19] This degree of chemical selectivity in photo-induced processes opens new perspectives for controlling reaction pathways.[35] Furthermore, we show that by exploiting the localization of hot-electron carriers we can selectively drive Au-S cleavage on individual strands of multi-element nanoantennas by just varying the light polarization and the wavelength. Similar to dip-pen lithography,[10] our selective functionalization platform can be rapidly adopted to incorporate different functional groups (bio-molecules, inorganic complexes, polymers) on an array of nanoparticles and to even position other nano-sized components, such as nanoparticles or quantum dots, on any targeted region. The unique characteristics of our desorption/re-adsorption strategy and the single-molecule super-resolution fluorescence imaging approach provides direct evidence on the nanoscopic mechanism of plasmon-induced molecular desorption under femtosecond laser irradiation. We found that Au-S desorption is induced by localized hot electrons and that thermal-driven reactions can be ruled out.[11,12] The tuneable anisotropic distribution of resistive losses provides a degree of surface-targeted specificity that could not be attained with thermal or electrical driven processes. Exploiting resistive losses at the sub-particle scale allows controlled and self-induced selective capping removal/replacement.

This can open interesting opportunities for the synthesis of novel nanomaterials, where tight control of these parameters dictates the final shape and size of the nano-objects.[36] Our approach unlocks the use of ultrafast, spatially anisotropic features of plasmon-induced hot electrons for positioning and selective functionalization at the nanoscale.

## Acknowledgements:

This work has been supported by the EPSRC through the Reactive Plasmonics Programme (EP/M013812/1), the Royal Society, and the Lee-Lucas Chair in Physics. E.C. acknowledge financial support from the European Commission through a Marie Curie fellowship.

## Author contributions:

S.S., Y.L. and E.C. conceived the idea and contributed to the design of the experiments. S.A.M. supervised the project. S.S. designed the imaging experiments, functionalized the gold nanostructures, performed the microscopy experiments (femtosecond irradiation and super-resolution imaging) and analyzed all the data. Y.L. designed and synthesized gold nanostructures using electron-beam lithography and conducted the SEM characterization. Y.L. also performed the numerical simulations and part of super-resolution imaging measurements. E.C. performed the dark-field spectroscopy experiments, conducted the electrochemical analysis of the functionalized gold nanostructures and performed the solution based-sorting experiments. All authors contributed to the general discussion and co-wrote the manuscript.

## Additional information:

**Supplementary information** accompanies this paper. Correspondence and requests for materials should be addressed to S.S.

**Competing financial interests:** The authors declare no competing financial interests.

# Supplementary Information

**Materials and Methods**

**Section 1. Materials**

Thiol-labelled and Atto 655 fluorescently modified single stranded DNA oligos were purchased from biomers.net GmbH (see Supplementary Table 1). Glass coverslips were purchased from VWR. Sticky-Slide VI 0.4 slides were purchased from Ibidi. Remover 1165 and poly(methyl methacrylate) (PMMA) 950K A4 were purchased from Dow and Microchem, respectively. Espacer 300Z was purchased from Showa Denko. 6-Mercapto-1-hexanol (MCH) and Tris (carboxyethyl) phosphine hydrochloride were purchased from Sigma-Aldrich (catalogue numbers: 451088 and C4706, respectively). Streptavidin functionalized gold nanoparticles were purchased from Nanopartz Inc. Self-assemble monolayer reagent for biotin terminate monolayer on gold surface was obtained from Sigma-Aldrich (catalogue number: 746622). Three buffers were used for sample functionalization and imaging: buffer A (10 mM TRIS-HCl, 1 M NaCl, pH = 7), buffer B (10 mM TRIS-HCl, 50 mM NaCl, pH = 7 and buffer C (5 mM TRIS-HCl, 10 mM $MgCl_2$, 1 mM EDTA, pH = 8).

**Section 2. Gold nanostructures fabrication**

Gold nanoantenna arrays with 3 µm pitch were fabricated on borosilicate glass coverslips using electron beam lithography (E-line Plus, Raith GmbH). The substrate was first coated with poly(methyl methacrylate) (PMMA) 950K A4 at 3,000 r.p.m. for 60 s and then baked for 3 min at 160 °C. Following the film preparation, a conductive layer of Espacer 300Z was spin coated onto the sample. Nanostructures were defined into PMMA by an electron beam exposure at 20 keV. After patterning, the spacer was removed using deionized water, followed by an MIBK/IPA (1:3) based development procedure. The pattern was then cleaned by a plasma ash etch step (Femto, Diener Electronic GmbH, 7 s at 40% power). Chromium and gold were coated at 1.5 Å/s for 1.5 nm and 30 nm via thermal evaporation (Amod, Angstrom Engineering Inc), and a standard lift-off process with Remover 1165 at 70 °C completed the fabrication process. The thickness of the chromium and gold layer was is 1.5 nm and 30 nm, respectively.

**Section 3. Scanning electron microscopy imaging**

To image the nanoantennas, 20 nm thin film of Espacer 300Z was coated onto the sample at 2,000 r.p.m. for 60 s, followed by a 60 s, 90 °C bake. For scanning electron microscopy imaging (E-line Plus, Raith GmbH) we typically used an electron-beam energy of 10 keV, a beam aperture of 30 µm and a working distance of 10 mm.

**Section 4. Dark-field spectroscopy**

Dark-field spectra of the scattering cross-sections of the antennas in air were measured with a diffractive grating and charge-coupled device camera cooled to 70 K. The plotted spectra are corrected for the system wavelength response (by measuring the cross section of a perfect reference white rough scatter) and also for dark and background counts. A dark field objective (Nikon LU Plan ELWD 100× NA 0.80) was used for both, to provide the incident illumination and to collect the scattered light.

**Section 5. Gold-sulfur functionalization.**

DNA functionalization of the previously fabricated gold nanoantenna arrays was performed as follows. First, the sample was thoroughly cleaned with isopropanol and water. Then, the sample was dried with nitrogen and left for 1.5 min in a UVO cleaner/oxidizer (Femto, Diener Electronic GmbH). The sample was mounted to a bottomless self-adhesive slide which, has been previously plasma clean for 10 min, to form a flow chamber of six individual channels with inner volumes of ~ 30 µL each. Each channel was functionalized independently from the other. First, the channel was rinse with 100 µL of buffer A. To reduce the disulfide bonds of the thiolated single stranded DNA (ssDNA) to monothiol, 20 µL of 100 µM modified oligo was incubated with 20 µL of 10 mM (Tris (carboxyethyl) phosphine hydrochloride for 30 minutes at room temperature. Then, a solution of activated 1 µM thiol-labelled ssDNA oligo diluted in buffer A was flown into the channel and incubated overnight in the dark and at room temperature. After washing the channel with 100 µL of buffer A, a solution of 1 mM 6-Mercapto-1-hexanol (MCH) was flown and incubated for 30 minutes. Finally, the sample was left in either buffer B or C (for femtosecond pulsed-laser irradiation or fluorescence imaging, respectively) after washing it with 100 µL of buffer B. For fluorescence microscopy imaging, a solution of the complementary imager strand diluted to the desired concentration in buffer C was flown into the channel immediately before starting the measurements. DNA re-functionalization of the gold nanoantennas was performed by flowing a 1 µM solution of another type of activated thiol-labelled single stranded DNA oligo in buffer A and incubating for 2 hs at room temperature. The backfilling step with MCH was performed exactly as described previously. Selective attachment of gold nanoparticles to gold nano-arrays, was performed by first incubating the gold nano-arrays with a 1 mM solution of MCH for 30 minutes. Following selective desorption, the sample was immediately immerse in a 1 mM biotin-SAM solution for 1 hs and clean by sequential washing steps of ethanol and water. The sample was finally washed with 1 mM PBS buffer and left to react with a solution of 50 nm streptavidin functionalized gold nanoparticles of OD =1 overnight. The solution of nanoparticles was removed and the sample was washed with 1 mM PBS buffer several times before SEM imaging.

**Section 6. Surface coverage and electrochemical measurements**

In order to estimate the surface coverage of the mix monolayer (MCH + thiolated-DNA docking strand) we performed reductive electrodesorption measurements on polycrystalline preferred oriented Au (111) electrodes. Self-assembly conditions were identical to the ones described in the gold-sulfur functionalization section. Measurements were performed in a CHI600E Potentiostat, IJ Cambria Scientific Ltd. Electrodesorption curves were recorded in a three-electrode glass cell equipped with a Pt foil as counter electrode and a Ag/AgCl reference electrode. The amount of thiolate species on the Au substrates was measured by reductive desorption in 0.1 M NaOH by a potential scan run from - 0.1 to -1.4 V at $v = 0.1$ V s$^{-1}$. The electrochemically-active surface area of the Au electrodes was voltammetrically estimated in the same electrolyte (0.1 M

NaOH solution) by recording a triangular potential sweep from -1.40 to 0.60 V at v = 0.1 V s$^{-1}$ and considering that the charge density related to the electroreduction of a gold oxide monolayer is 0.44 mC cm$^{-2}$. Typical reductive desorption curves for the mix SAMs from the Au substrates are shown in Supplementary Figure 1. The cathodic peak correspond to the reductive desorption of the thiolate species from the Au surface. The charge density involved in these peaks gives an accurate estimation of the amount of chemisorbed thiolate to the Au surface. Integration of the voltammetric peaks show a value of q ≈ 80±5 μC cm$^{-2}$, that corresponds to a thiolate coverage θ ≈ 1/3, assuming one electron transfer for the reductive desorption reaction. The value θ = 1/3 is the maximum surface coverage reached by thiolate species on Au surfaces.[1] From SEM images we estimated the surface area of a single leg (A ≈ 2.28 x 10$^4$ nm$^2$) in the nanotrimer antenna design. Using the obtained value for the surface coverage on the Au (111) electrode, we estimate the number of Au-S bonds per leg of the nanotrimer antenna design to be 1.05 x 10$^5$ molecules in the mix monolayer. From these value and using reported data[2] on the coverage of thiolated ss-DNA under similar experimental conditions, we determined a ratio of ~ 160/1 of MCH/ss-DNA molecules in our mix monolayer.

**Section 7. Femtosecond pulsed-laser irradiation.**

Selective desorption of thiol-modified oligos from gold nanoantennas was performed in a home-built optical set up which is based on an Nikon Eclipse Ti-U microscope (Nikon Instruments) that combines raster scanning of a sample with femto-second pulsed laser irradiation, together with, white light illumination and detection using a CCD camera (QICAM Fast 1394, QImaging). This configuration enables the automation and live monitoring of the single nanoantennas irradiation. Femtosecond pulsed-laser irradiation of individual nanoantennas was carried out using a collinear optical parametric amplifier (ORPHEUS, Light Conversion Ltd.) pumped by a femtosecond Ytterbium based laser system (PHAROS, Light Conversion Ltd.) operating at 100 kHz and with a pulse duration of ~ 180 fs. The excitation wavelength was set either to 820, 950 or 1100 nm and the beam polarization was adjusted with a half-wave plate and a polarizer. The laser was coupled into the microscope objective (CFI S Plan Fluor ELWD 40x, NA 0.60, air) with a short-pass dichroic mirror of either 805 (DNSP805, Thorlabs) or 1000 nm (DMSP1000, Thorlabs), respectively, and focused onto the sample. The sample was fixed to an XYZ piezo-scanner stage (Nano-Drive, Mad City Laboratories) to perform the scanning. Each nanoantenna was irradiated for 10 seconds at power densities from 9 to 526 W/cm$^2$. Following irradiation a washing step consisting of at least three washes using 100 μL of buffer B was performed.

**Section 8. Fluorescence microscopy imaging**

Fluorescence imaging was carried out on a custom-built total internal reflection fluorescence microscope, based on an inverted Nikon Eclipse Ti-U microscope (Nikon Instruments) and equipped with an oil-immersion TIRF objective (Apo DIC N2 TIRF 100x, numerical aperture (NA) 1.49, oil). The effective pixel size is of 160 nm. Fluorescence excitation was at 640 nm using a 20 mW CW laser (LDH, Picoquant) spectrally filtered with a clean-up filter (BrightLine HC 636/8, Semrock). The laser was coupled into the microscope objective using a single-edge dichroic beam splitter (BrightLine FF 649, Semrock) and focused on the back focal plane of the objective aligned for TIRF illumination. Fluorescence light was spectrally filtered using an emission filter (ET705/72m, Chroma Technology) and imaged on an electron-multiplying charge-coupled device (EMCCD) camera (Evolve 512 Delta, Photometrics). For DNA-PAINT measurements, 20,000 frames per super-resolution image were recorded at a frame rate of 10Hz and at an excitation intensity of ~ 450 W cm$^{-2}$. The CCD readout bandwidth was set to 5 MHz at 16 bit and 3 pre-amp gain. The electron multiplying (EM) gain was set at 200. Imaging conditions were as follows. The concentration of the imager strand was determined empirically to guarantee biding of only one single imager strand per diffraction limited-area.

**Section 9. Super-resolution data processing and image analysis**

Super-resolution images were reconstructed from the raw data based on subsequent localization of single molecules via sparse segmentation and Gaussian fitting using the rainSTORM Matlab application.[3] Prior single molecule localization, stage drift correction and background subtraction of the photoluminescence (static, auto-fluorescence emission) from the gold nanoantennas were performed with an in-house Matlab routine. Specifically, an efficient subpixel image registration algorithm based on cross-correlation was used for drift correction to within 1/10 of a pixel precision. This algorithm is referred to as the single-step DFT algorithm in Supplementary Reference 4.[4] Background subtraction was performed using the average image of the drift-corrected image stack, which reduces the static background gold luminescence without masking the detection of the fluorescent single molecule binding events. Images of individual gold nanoantennas were typically reconstructed from 700 to 3,000 localizations. To reconstruct high-density super-resolution images of gold nanoantennas exposed to specific chemical and/or optical conditions, we overlaid super-resolution images of 28 to 36 individual gold nanoantennas that had been exposed to the same conditions. The alignment of the individual nanoantennas was possible by calculating the coordinates of the centre of each nanoantenna with high accuracy. This was performed using a translation registration algorithm (Matlab built-in function), which estimates the geometric transformation that aligns an image with a fixed template. The template was design according to the shape and dimension of each type of fabricated nanoantenna.

**Section 10. Solution experiments for capping displacement**

Citrate-capped 50 nm Au spherical nanoparticles OD =1 (AuNPs) and 150 x 50 nm Au nanorods OD=1 (AuNRs) were modified with 1 mM 4-mercaptobenzoic acid (MBA) aqueous solution for 2 hours. The nanoparticles were later purified by two centrifugation steps at 4,000 rpm for 10 minutes and re-dispersed in water. Illumination of the colloidal mix (1:1 ratio) was performed in a quartz cuvette (3 ml total volume) under magnetic stirring conditions. Samples were irradiated at 780 nm for 10 minutes at 10 mW average power using the femtosecond laser previously described. Further increase on the irradiation time or power leads to AuNRs deformation.[5] For SEM imaging quartz substrates were modified with APTES (5% in ethanol) for 1 hour followed by vigorous rising with ethanol. The solution containing the irradiated and non-irradiated nanoparticles were left overnight in a home-made humidity chamber. Many washing steps with water and final nitrogen drying were performed before SEM imaging.

**Section 11. Theoretical simulations**

Theoretical calculations of the scattering/absorption cross-section spectra and the near-field intensity distributions were performed using both Lumerical FDTD 2017a and Comsol Multiphysics v4.4. The photo-thermal heating of gold nanoantennas for femtosecond laser irradiation was studied by coupling the Radio Frequency module with the Heat Transfer module in Comsol. A dual-hyperbolic two-temperature model was adapted to study optical losses and their decay into the surrounding media (see Section 12). The temperature dependent parameters of gold, glass, and water were taken from the Comsol material library.

**Section 12. Dual-hyperbolic two-temperature model**

Hot electrons excited upon resonant surface plasmons decay collide with electrons around the Fermi level. These hot electrons are then diffused into deeper parts of the electron gas at a speed (<$10^4$ m/s) much lower than that of the ballistic motion (close to the Fermi velocity ~ $10^6$ m/s) while transferring their energy to the lattice through electron-phonon coupling.[6] The non-equilibrium between electrons and lattice has been observed experimentally[7,8] and can be described by a well-established two-temperature model, which was originally proposed by Anisimov et al.[9]

We simulate the temporal dependence of the lattice temperature of gold nanostructures supported on a silica substrate and embedded in water by using the hyperbolic model to calculate the heat conduction, in both, electrons and phonons after femtosecond irradiation. The energy equation for the lattice becomes,

$$C_e \frac{\partial T_e}{\partial t} + C_e \tau_e \frac{\partial^2 T_e}{\partial t^2} = \kappa_e \nabla^2 T_e - G(T_e - T_l) - \tau_e \frac{\partial}{\partial t} G(T_e - T_l) + Q + \tau_e \frac{\partial Q}{\partial t}$$

$$C_l \frac{\partial T_l}{\partial t} + C_l \tau_l \frac{\partial^2 T_l}{\partial t^2} = \kappa_l \nabla^2 T_l + G(T_e - T_l) + \tau_l \frac{\partial}{\partial t} G(T_e - T_l)$$

where $T_e$ and $T_l$ are the electron and lattice temperature, $\tau_e$ and $\tau_l$ are the electron and lattice collision time of 0.04 ps and 0.8 ps for gold,[10] respectively; $\kappa_e$ and $\kappa_l$ are the electronic and lattice thermal conductivity of gold; $C_e$ and $C_l$ are the electronic and lattice heat capacity and where $\kappa_e = \frac{T_e}{T_l}\kappa_{e,0}$ and $\kappa_{e,0} \approx \kappa_l$. Within the free electron gas model[11,12] $C_e = \frac{\pi^2 k_B^2 n_e}{2E_F} T_e$ and can be associated to the free electron number density $n_e$. We use a gold-water extraction potential $E_F$ of 3.72 eV.[13] Since the Debye temperature for gold is 178 K at room temperature,[14] the electron-phonon coupling $G$ for gold can be assumed to be independent to the electron temperature as 2.6×$10^{16}$ W/($m^3$ K). $Q$ represents the resistive loss of gold nanoantenna from optical absorption.

# Supplementary Tables

| Description | Sequence |
|---|---|
| Docking $P_1$ | ThiolC6-SpacerC18-ATTACTTCTTT |
| Docking $P_2$ | ThiolC6-SpacerC18-ATGAGTTAATT |
| Imager $P_1^*$ | Atto655-AGAAGTAATG-3' |
| Imager $P_2^*$ | Atto655-TTAACTCATG-3' |

**Supplementary Table 1**. Metallic DNA-PAINT docking and imager sequences

**Supplementary Figures**

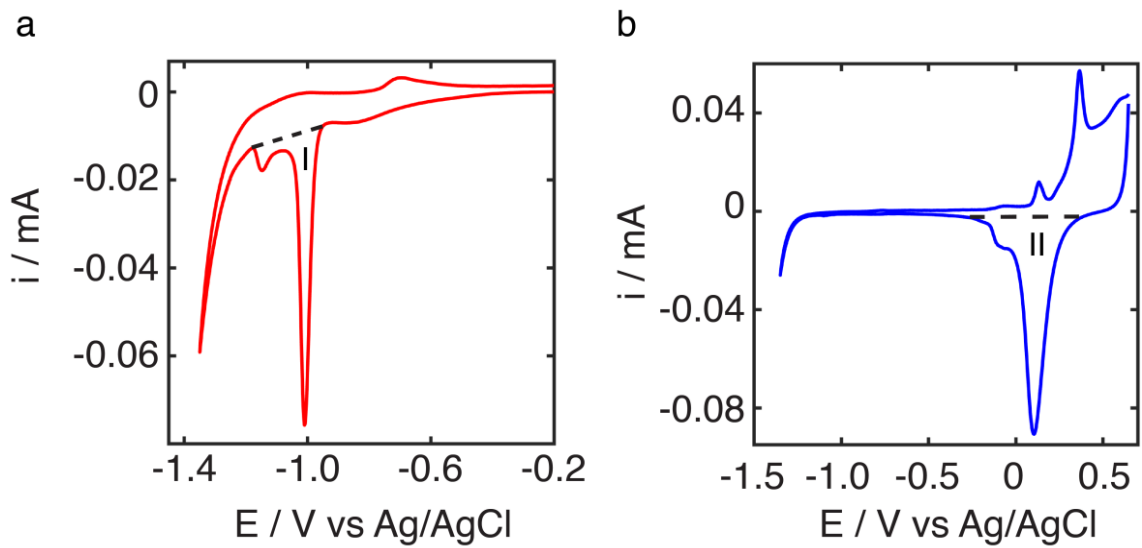

**Supplementary Figure 1. Surface coverage estimation from electrochemical measurements. a,** Reductive electrodesorption curve for the mix monolayer (MCH + ss-DNA) SAM on a Au (111) substrate. Integration of the charge in peak I indicates the amount of Au-S bonds cleaved. **b,** Electrochemically active surface area of the Au electrode was determined after thiols desorption. Integration of the charge from peak II allows to estimate the charge density and surface coverage, as explained in Section 6.

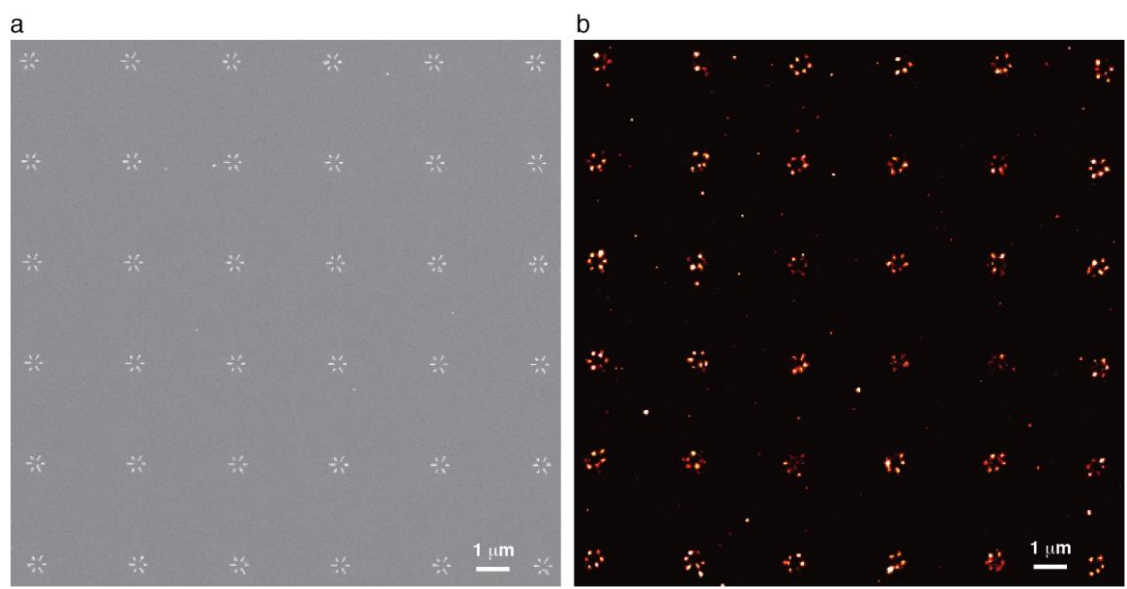

**Supplementary Figure 2. Super resolution m-PAINT imaging of gold nanostructures**. **a**, SEM and **b**, super-resolved metallic DNA-PAINT image of a 6 x 6 array of individual six-leg nano-sized gold structures fabricated with electron beam lithography. m-PAINT movie was collected using 0.1 nM concentration of imager strand $P_1^*$ in imaging buffer C. The mean number of localizations per individual nanostructure is 1,458 ± 332.

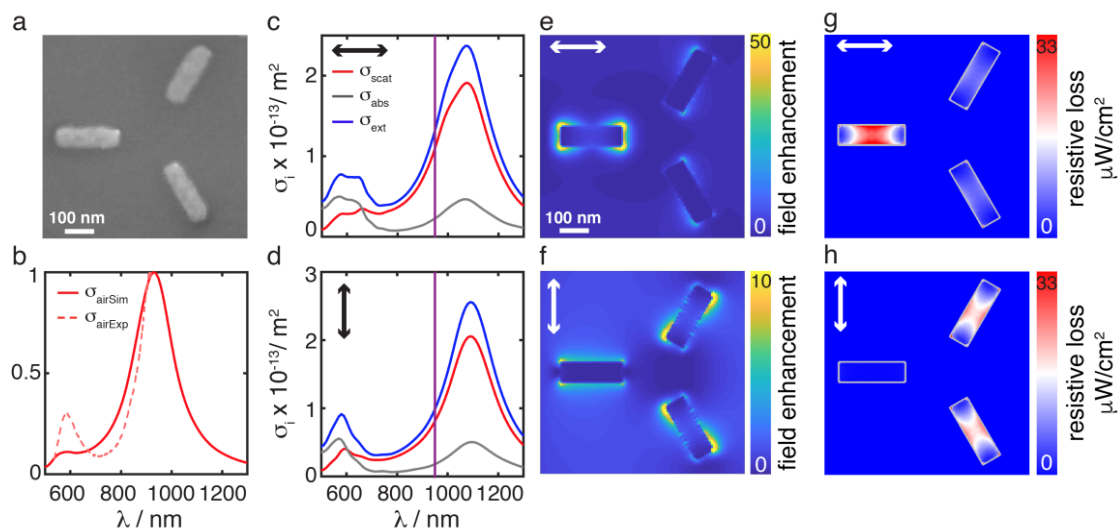

**Supplementary Figure 3. Plasmonic response of gold nanotrimer structures**. **a**, SEM image of a gold trimer nano-sized structure composed of three identical nanorods with dimension of 180 nm x 55 nm. **b,** FDTD-simulated (full line) and measured (dashed line) single-antenna scattering spectra in air. **c, d,** Simulated scattering (red), absorption (grey) and extinction (blue) spectra in water for parallel and perpendicular polarized illumination, respectively. Violet lines highlight the laser wavelength used in the desorption experiments (950 nm). **e, f,** FDTD simulations of the near-field distribution at 950 nm in water for parallel and perpendicular polarization, respectively. Colour scale bars represent the field enhancement ($|\mathbf{E}|/|\mathbf{E}_0|)^2$ values. **g, h,** Comsol simulations of the resistive loss distribution at 950 nm after femtosecond laser irradiation in water (fluence of 1.7 mJ/cm$^2$) for parallel and perpendicular incident polarization, respectively. Colour scale bars represent the resistive losses values in µW/nm$^3$.

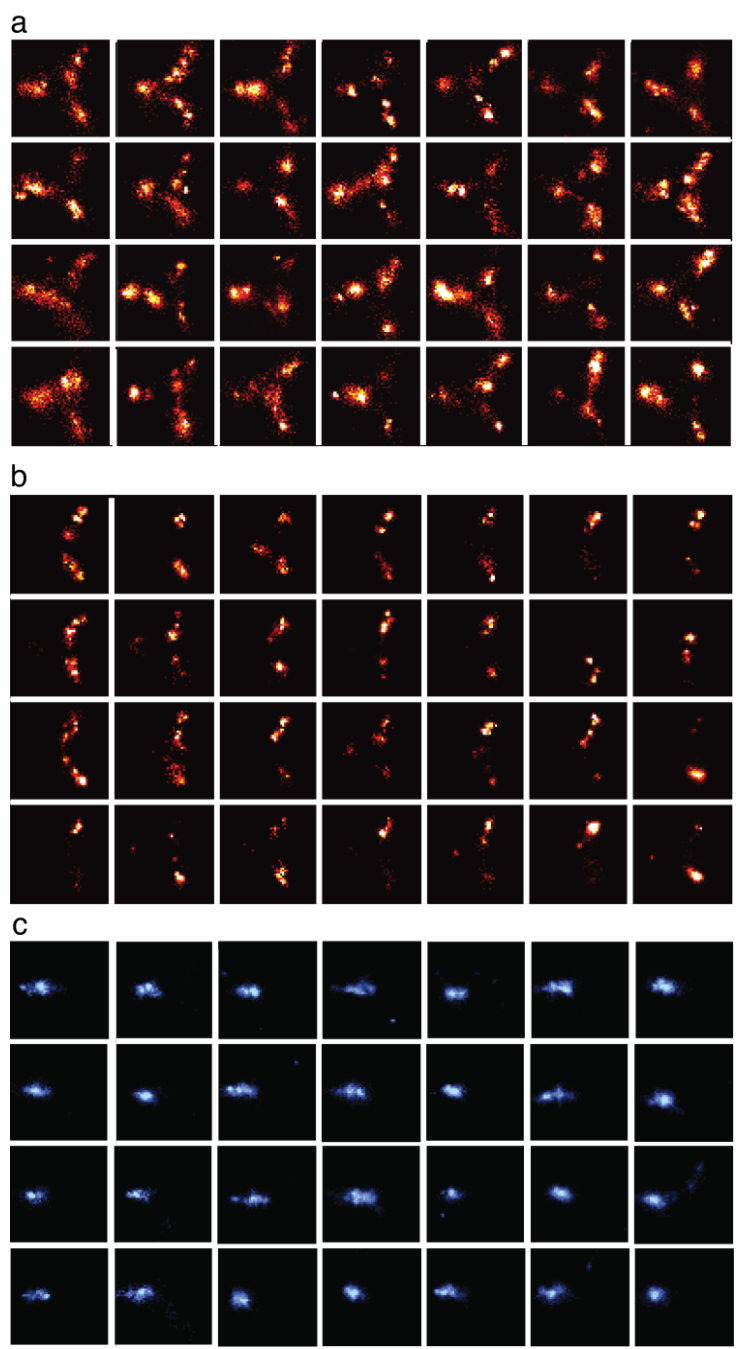

**Supplementary Figure 4. Super-resolution m-PAINT imaging of thiol-desorption and ligand replacement from individual gold nanotrimer structures.** Cropped super-resolved metallic DNA-PAINT images of individual gold trimer nano-sized structures used to generate the high-density super-resolved images (see Figure 1c) before (**a**) and after (**b**, **c**) femtosecond laser irradiation (950 nm, 1.7 mJ/cm$^2$) and re-functionalization (**c**) with the thiol-reactive docking strand P$_2$. The polarization of the light was aligned parallel to the horizontal leg of the trimer antenna. m-PAINT movie was collected using 1 nM or 2 nM concentration of imager strand P$_1$* (**a**, **b**) or P$_2$* (**c**) in imaging buffer C, respectively. Pseudo colors red and blue represent docking strands P$_1$ and P$_2$, respectively.

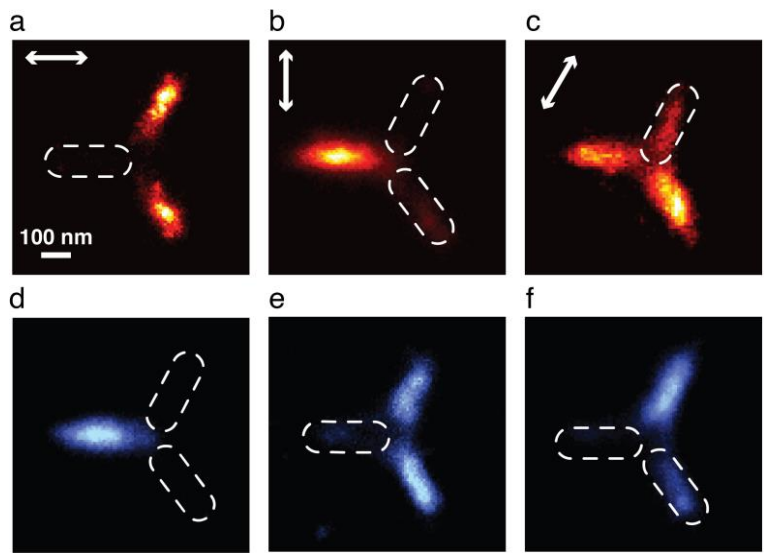

**Supplementary Figure 5. Controlling thiol-desorption and ligand replacement from gold nanotrimer structures with polarized light illumination.** Average super-resolved metallic DNA-PAINT images of *n* individual gold nanotrimers carrying two DNA docking strand species (pseudo colors red and blue for docking strands $P_1$ and $P_2$, respectively). The dual functionalization of the nanostructures was achieved by first, irradiating the antenna with a linearly polarized femtosecond pulsed laser (950 nm, 1.7 mJ/cm$^2$) at 0° (**a** and **d**), 90° (**b** and **e**) or 120° (**c** and **f**) and then, by re-functionalization with the thiol-reactive docking strand $P_2$. Imaging of individual targets using the same fluorophore was performed by a round of acquisition with imager strand $P_1$*, followed by a washing step, and the sequential incubation of imager strand $P_2$*. m-PAINT movies were acquired using a 1 or 2 nM concentration of imager strand $P_1$* (**a**, **b**, **c**) or $P_2$* (**d**, **e**, **f**) in imaging buffer C, respectively. *n* and the mean number of localizations in individual images is 28 and 776 ± 480, 34 and 1,348 ± 545, 34 and 1,946 ± 552, 28 and 1,080 ± 597, 34 and 1,426 ± 721, 34 and 2,338 ± 1,019 for **a**, **b**, **c**, **d**, **e**, **f**, respectively.

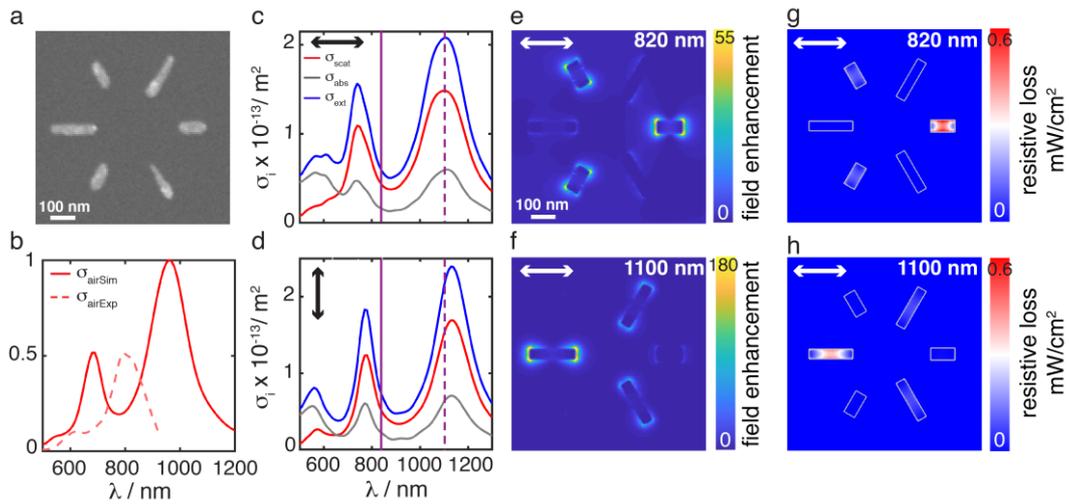

**Supplementary Figure 6. Plasmonic response of six-leg gold nanostructures**. **a,** SEM image of a six-leg gold nanostructure composed of two sizes of nanorods with dimensions of 110 nm x 45 nm and 180 nm x 40 nm. **b,** FDTD-simulated (full line) and measured (dashed line) single-antenna scattering spectra in air. **c, d,** Simulated scattering (red), absorption (grey) and extinction (blue) spectra in water for parallel and perpendicular polarized illumination, respectively. Violet lines highlight the laser wavelength used in the desorption experiments (820 or 1100 nm). **e, f,** FDTD simulations of the near-field distribution at 820 nm and at 1100 nm in water for parallel polarization, respectively. Color scale bars represent the field enhancement $(|E|/|E_0|)^2$ values. **g, h,** Comsol simulations of the resistive loss distribution at 820 nm or at 1100 nm after femtosecond laser irradiation of 1.9 mJ/cm² or 3.8 mJ/cm² in water for parallel incident polarization, respectively. Color scale bars represent the resistive losses values in mW/nm³.

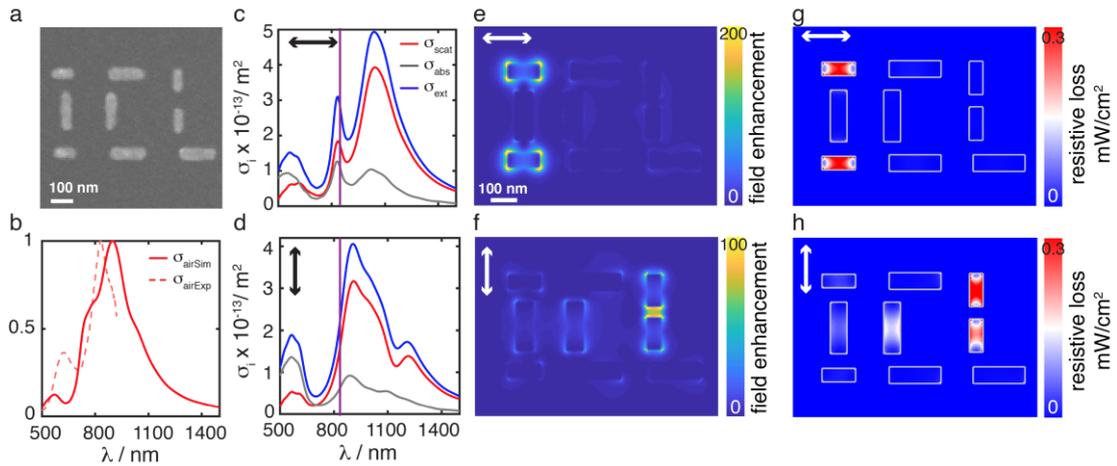

**Supplementary Figure 7. Plasmonic response of 'ICL' gold nanostructures**. **a,** SEM image of a gold custom-design letter pattern ('ICL') nanostructure composed of two sizes of nanorods with dimensions of 110 nm x 45 nm and 170 nm x 50 nm. **b,** FDTD-simulated (full line) and measured (dashed line) single-antenna scattering spectra in air. **c, d,** Simulated scattering (red), absorption (grey) and extinction (blue) spectra in water for parallel and perpendicular polarized illumination, respectively. Violet lines highlight the laser wavelength used in the desorption experiments (820 nm). **e, f,** FDTD simulations of the near-field distribution at 820 nm in water for parallel and perpendicular polarization, respectively. Color scale bars represent the field enhancement ($|E|/|E_0|)^2$ values. **g, h,** Comsol simulations of the resistive loss distribution at 820 nm after femtosecond laser irradiation of 2.6 mJ/cm$^2$ in water for parallel and perpendicular incident polarization, respectively. Color scale bars represent the resistive losses values in mW/nm$^3$.

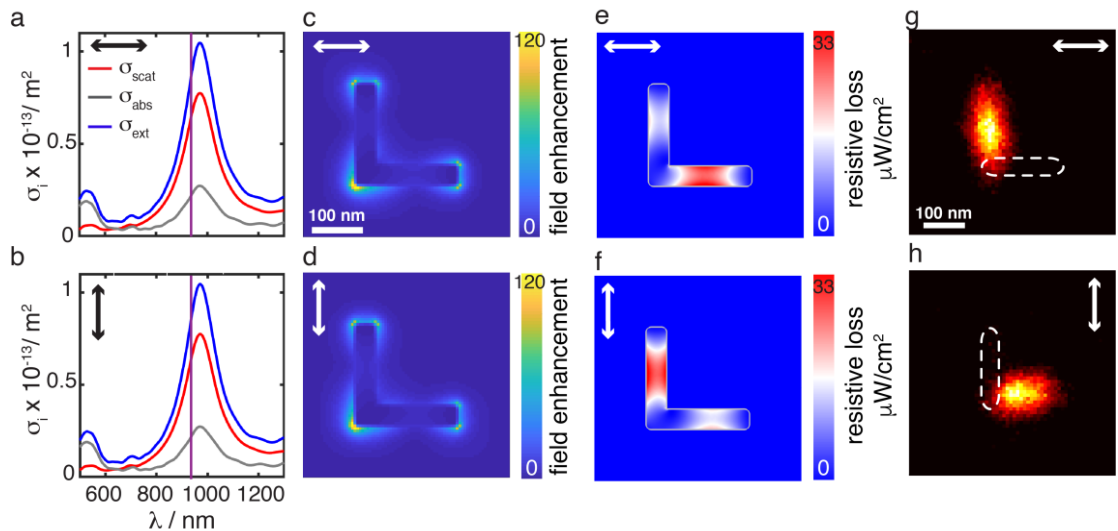

**Supplementary Figure 8. Plasmonic response of 'L' gold nanostructures**. **a, b,** Simulated scattering (red), absorption (grey) and extinction (blue) spectra of a L-shape gold nanostructure composed of two connected nanorods, with dimensions of 200 nm x 40 nm, in water for parallel and perpendicular polarized illumination, respectively. Violet line highlights the laser wavelength used in the desorption experiments (950 nm). **c, d,** FDTD simulations of the near-field distribution at 950 nm in water for parallel and perpendicular polarization, respectively. Color scale bars represent the field enhancement $(|E|/|E_0|)^2$ values. **e, f,** Comsol simulations of the resistive loss distribution at 950 nm after femtosecond laser irradiation of 5.3 mJ/cm$^2$ in water for parallel and perpendicular incident polarization, respectively. Color scale bars represent the resistive losses values in μW/nm$^3$. **g, h,** Average super-resolved metallic DNA-PAINT images of $n$ individual gold nanostructures functionalized with docking strand P$_1$ after irradiation with a 950 nm linearly polarized femtosecond pulsed laser of 5.3 mJ/cm$^2$ fluence. White arrows represent the polarization of the incident light. m-PAINT movies were collected using 5 nM of imager strand P$_1$*. $n$ and the mean number of localizations in individual images is 33 and 701 ± 436, 30 and 634 ± 448, respectively.

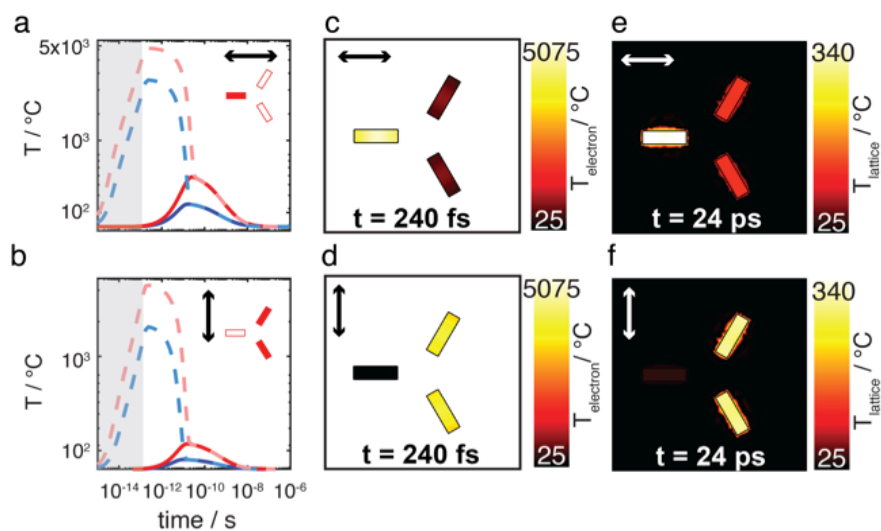

**Supplementary Figure 9. Numerical calculation of temperature profiles of plasmon-selective Au-S desorption on gold nanotrimers. a, b,** Simulated temporal evolution of the electron temperature (dashed line) and lattice temperature (full line) for the horizontal and vertical elements of a gold nanotrimer structure composed of three identical nanorods with dimension of 180 nm x 55 nm following femtosecond laser irradiation (0°, 950 nm) at 0.6 mJ/cm² (blue) and 2.0 mJ/cm² (red) fluence. **c-f,** Map distribution of the electron temperature (**c**, **d**) and the lattice temperature (**e**, **f**) following femtosecond laser irradiation at 240 fs and 24 ps delayed after the temporal start of a 180 fs laser pulse. Black and white arrows represent the polarization of the incident light.
27

# Supplementary References